\documentclass[ 3p, 10pt, twocolumn]{elsarticle}

\usepackage[utf8]{inputenc}
\usepackage{float}
\usepackage[pdfversion=1.4]{hyperref}
\usepackage{mathtools}
\usepackage{siunitx}
\DeclareSIUnit{\au}{a.u.}

\usepackage{physics}
\usepackage{amsfonts}
\usepackage{bm}
\usepackage{enumitem}
\usepackage{subcaption}
\setlist[description]{leftmargin=1.5\parindent, labelindent=\parindent,
rightmargin=1.5\parindent}
\usepackage{tikz}
\usetikzlibrary{arrows,automata}
\usetikzlibrary{positioning}

\newcommand{\pkg}[1]{\texttt{#1}}
\newcommand{\fiend}{\pkg{Fiend}}
\newcommand{\fenics}{\pkg{FEniCS}}
\newcommand{\iu}{{\mathrm{i}\mkern1mu}}
\newcommand{\D}{\mathrm{d}}

\biboptions{sort&compress,numbers,square,comma}

\journal{Computer Physics Communications}

\begin{document}

\begin{frontmatter}

    \title{\fiend\ -- Finite Element Quantum Dynamics}

    \author{J. Solanpää}
    \cortext[a]{janne@solanpaa.fi}
    \author{E. Räsänen}
    \address{Computational Physics Laboratory, Tampere University, Tampere FI-33101, Finland}

    \begin{abstract}
We present \fiend\ -- a simulation package for three-dimensional single-particle time-dependent Schrödinger
equation for cylindrically symmetric systems. \fiend\ has been designed for the simulation
of electron dynamics under \emph{inhomogeneous} vector potentials such as in nanostructures, but it
can also be used to study, e.g., nonlinear light-matter interaction in atoms and linear molecules.
The light-matter interaction can be included via the minimal coupling principle in its full rigour, beyond the
conventional dipole approximation. The underlying spatial discretization is based on the finite element method (FEM),
and time-stepping is provided either via the generalized-$\alpha$ or Crank-Nicolson methods. The software is
written in Python 3.6, and it utilizes state-of-the-art linear algebra and FEM backends for performance-critical
tasks. \fiend\ comes along with an extensive API documentation, a user guide, simulation examples,
and allows for easy installation via \pkg{Docker} or the Python Package Index.
\end{abstract}


\end{frontmatter}

\clearpage
\onecolumn

{\noindent\bf PROGRAM SUMMARY}\\
\begin{small}
\noindent
{\em Program Title:} \fiend\  \\
{\em Journal Reference:}                                      \\
{\em Catalogue identifier:}                                   \\
{\em Licensing provisions:} MIT License \\
{\em Programming language:} Python 3.6 \\
{\em Computer:}  Tested on x86\_64 architecture.    \\
{\em Operating system:}  Tested on Linux and macOS. \\
{\em RAM:} Simulation dependent: from megabytes to hundreds of gigabytes. \\
{\em Parallelization:} MPI-based parallelization. Thread-based parallelization via
\pkg{BLAS} and \pkg{LAPACK} backends. \\
{\em Classification:}
2.2 Spectra, 2.5 Photon Interactions, 4.3 Differential equations, 6.5 Software including Parallel Algorithms \\
{\em External routines/libraries:}                            \\
\pkg{PETSc}, \pkg{SLEPc}, \pkg{FEniCS}, \pkg{HDF5} \\
\pkg{petsc4py}, \pkg{slepc4py}, \pkg{mpi4py}, \pkg{h5py}, \pkg{numpy}, \pkg{scipy}, \pkg{matplotlib}, \pkg{psutil}, \pkg{mypy}, \pkg{progressbar2} \\
{\em Nature of problem:}\\
Solution of time independent and time dependent single active electron Schrödinger equations
in cylindrically symmetric systems including interactions with spatially inhomogeneous vector potentials.  \\
{\em Solution method:}\\
Finite element discretization of the Schrödinger equation. Time evolution
via the generalized-$\alpha$ or Crank-Nicolson methods.  \\
{\em Restrictions:}\\
Cylindrically symmetric single active electron systems.\\
{\em Unusual features:}\\
Finite element discretization of the equations allowing \emph{inhomogeneous spatial dependence of
the vector potential} (e.g., plasmon-enhanced fields). Integration with the FEniCS finite element suite.
Installable from the Python Package Index. Pre-installed \pkg{Docker} images available.\\
{\em Additional comments:}\\
The source code is also available at \url{https://fiend.solanpaa.fi}. \\
Python package available at \url{https://pypi.org/project/fiend/}.\\
Docker images are available at \url{https://hub.docker.com/r/solanpaa/fiend/}.\\
{\em Running time:}\\
From minutes to weeks, depending on the simulation.
\end{small}
\clearpage
\twocolumn

\section{Introduction}

\noindent Simulation of three-dimensional (3D) single particle quantum mechanics (QM) is still one of
the most used computational approaches in the strong field and attosecond communities.
In these fields, the nonlinear interaction of the atomic or molecular electron with the driving laser field
requires fast and accurate integration of the 3D time dependent Schrödinger equation (TDSE).
Recent applications
include, e.g., benchmarking of approximate models~\cite{PhysRevA.95.033415,PhysRevA.94.013415,PhysRevA.93.023422,PhysRevA.95.023403,0953-4075-49-5-053001,PhysRevA.98.023401},
study of high-order harmonic generation~\cite{1674-1056-27-7-073205,Abdelrahman:18,PhysRevLett.113.033001},
and the study of photoionization~\cite{PhysRevA.90.043406,0953-4075-48-2-025601,Sayler:15,0953-4075-51-10-104003,1367-2630-20-6-063018}.

Recently the strong-field and attosecond communities have turned their attention to related
phenomena in nanostructures. There the nanostructure geometry and plasmonic effects
cause the electromagnetic field to become extremely inhomogeneous~\cite{nanoplasmonic_review} which, in turn,
causes significant differences to the traditional ultrafast and strong-field phenomena
in atoms and molecules. Recent studies include, e.g., nanostructure-enhanced
photoionization of gases~\cite{PhysRevLett.119.053204,PhysRevA.97.023420,Ciappina:18} and
electron emission from nanostructures such as tips~\cite{Piglosiewicz2013,kruger2012,Wimmer2014,doi:10.1063/1.4934681,doi:10.1063/1.4991681,1367-2630-9-5-142,herink2012,PhysRevLett.105.147601,Forg2016,field_local_resc}
and rods~\cite{0957-4484-25-46-465304,doi:10.1063/1.4927151,doi:10.1021/nl302271u,doi:10.1021/nn504594g,doi:10.1021/nn305194n}.

There are already multiple software designed for integrating the
three-dimensional time-dependent Schrödinger equation (TDSE).
Open-source software include, e.g.,
Qprop~\cite{qprop}, Octopus~\cite{octopus1,octopus2}, QnDynCUDA~\cite{qndyncuda}, and
WavePacket~\cite{wavepacket}, but there are also plenty of other options
such as SCID-TDSE~\cite{scidtdse}, tRecX~\cite{PhysRevA.81.053845,Tao_2012}, and ALTDSE~\cite{altdse}.
Qprop, SCID-TDSE, and QnDynCUDA solve the TDSE using a grid-based representation of the radial coordinates and spherical harmonics for the angular dependence;
TRecX and WavePacket support multiple basis sets; Octopus relies on a real-space grid; and ALTDSE requires end-user to provide the matrices in an appropriate eigenbasis.

However, efficient and accurate simulation of ultrafast and strong-field phenomena in nanostructures and nanostructure-enhanced
gases requires specialized TDSE-solvers. Most importantly, since the laser electric field
has strong spatial inhomogeneity, it is imperative to have position-dependent control of simulation accuracy.
In addition, it would be beneficial to have
the TDSE-solver directly interface with a solver for the electromagnetic problem.

To this end, we have developed \fiend, a QM simulator based on finite element method (FEM).
\fiend\ provides solvers for the time independent and time-dependent single active electron Schrödinger
equations in cylindrically symmetric systems\footnote{Although the restriction to cylindrically symmetric systems can be lifted relatively easily if needed.}
for solving time-independent and time-dependent Schrödinger equations in cylindrically symmetric systems.
We have designed \fiend\ to integrate with the open source \fenics\ finite element suite~\cite{fenics1,fenics2,dolfin1,ffc1,ffc2,ufl1,fiat1}
allowing for easy description of complicated system geometries. Being FEM-based, \fiend\ also allows
the use of potentials with integrable singularities.

\fiend\ is written in \texttt{Python 3.6},
but much of the heavy number-crunching is delegated to well-tested and efficient external libraries.
The software is modular and easy to extend. We also provide a comprehensive unit- and integration-test suite
to ensure reliability. Moreover, we provide straightforward installation methods
either via the Python Package Index~\cite{fiend_pypi} or as Docker images~\cite{docker}.

This paper is organized as follows. In Sec.~\ref{sec:theory} we describe the
finite element discretization of the Schrödinger equations and the time-propagation
schemes. Section~\ref{sec:implementation} focuses on the design and implementation of \fiend.
Section~\ref{sec:demos} provides a few example simulation, and finally, in Sec.~\ref{sec:summary}
we summarize this paper and possible extensions of the presented software suite.

\section{Finite element quantum mechanics}
\label{sec:theory}

\subsection{Systems}
\noindent\fiend\ is designed for simulating cylindrically symmetric
single (active) electron systems interacting with time
and (optionally) position dependent vector potentials.
The Hamiltonian operator of these systems can be written as
\begin{equation}
\begin{split}
    \hat{H} &= \hat{T_\rho} + \frac{ \hat{p}_z^2}{2}
    + V(\hat{\rho}, \hat{z}) + W(\hat{\rho}, \hat{z}, \hat{p}_\rho, \hat{p}_z, t),
\end{split}
\end{equation}
where $\hat{\rho}$ is the planar radial coordinate operator, $\hat{z}$ the $z$-coordinate operator,
$\hat{T}_\rho$ the kinetic energy operator with respect to planar radial motion\footnote{
Note that $\hat{T}_\rho$ is not proportional to the square of the planar radial momentum operator, i.e.,
\begin{equation}
\begin{split}
 \hat{T}_\rho &= -\frac{1}{2}\left(\frac{\partial^2}{\partial_\rho^2}+\frac{1}{2}\frac{\!\partial}{\partial_\rho}\right)\\
 &\neq\frac{1}{2}\left[-\iu\left(\frac{\!\partial}{\partial_\rho}+\frac{1}{2\rho}\right)\right]^2 = \frac{\hat{p}_\rho^2}{2}.
\end{split}
\end{equation}
This is a peculiarity of quantum mechanics in cylindrical coordinates~\cite{radial_momentum_operator}.},
$\hat{p}_{z}$ the $z$-component of the momentum operator,
$V$ the static potential, and $W$ the light-matter interaction operator.

\fiend\ readily supports three types of interaction operators $W$. First,
for linearly polarized vector potentials $\mathbf{A} = f(t) \bm e_z$
in the length gauge the interaction operator is given by
\begin{equation}
        W_{\mathrm{LG}}=-\frac{\partial f(t)}{\partial t} \hat{z},
\end{equation}
secondly, in the velocity gauge the interaction operator reads
\begin{equation}
        W_{\mathrm{VG}}= f(t) \hat{p}_z,
\end{equation}
and finally for more general cylindrically symmetric inhomogeneous vector potentials
$\mathbf A = \mathbf{A}_s(\rho, z) f(t)$ we provide the interaction operator
\begin{equation}
\begin{split}
W_{\mathrm{inhomogeneous}} &= \frac{1}{2}f(t) \mathbf{A}_s \cdot \mathbf{\hat{p}}+\frac{1}{2}f(t) \mathbf{\hat{p}} \cdot\mathbf{A}_s \\
                &+ \frac{1}{2}f(t)^2 \Vert \mathbf A_s(\hat{\rho}, \hat{z})\Vert^2.
\end{split}
\end{equation}
Moreover, \fiend\ supports
all interaction operators of the form $\hat{W} = W(\hat{\rho}, \hat{z}, \hat{p}_\rho, \hat{p}_z, t)$.

\begin{figure}
\centering
    \begin{subfigure}{0.5\linewidth}
        \centering
        \begin{tikzpicture}
            \tikzstyle{ann} = [font=\small,inner sep=1pt]

            \filldraw[fill=blue!20, fill opacity = 0.8, draw=none]
                (1.26, 2.35) arc ( 30.0:-90.0:1.48cm and 2.97cm)
                --  (0, 2.35) -- cycle;
            \draw[orange, line width = 1pt] (1.26,2.35) -- (0, 2.35) ;
            \draw[blue, line width = 1pt] (1.26, 2.35) arc ( 30.0:-90.0:1.48cm
            and 2.97cm);

            \draw[arrows=->] (0,0) to (2,0);
            \draw[arrows=->] (0,0) to (0,2.75);
            \node[ann] at (-0.2, 2.55) {$z$};
            \node[ann] at (1.8, -0.2) {$\rho$};

            \node[inner sep=0pt] (mesh) at (0.48,0)
                {\phantom{\includegraphics[width=0.65\textwidth]{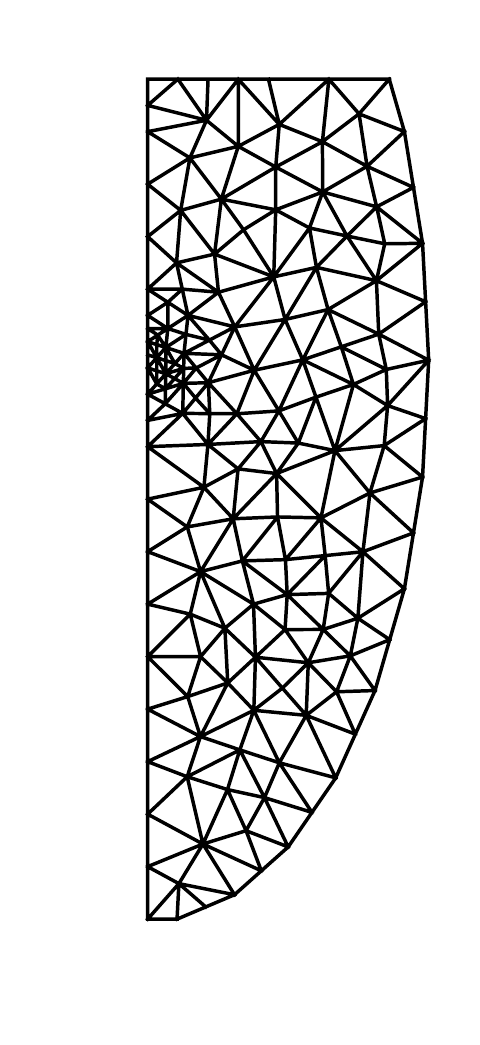}}};

            \node[ann] at (0.63, 2.55) {$\Gamma_N$};
            \node[ann] at (1.75, 0.5) {$\Gamma_D$};
            \node[ann] at (0.7, 0.8) {$\Omega$};
            \node[ann] at (-0.5, 2.85) {(a)};
        \end{tikzpicture}
    \end{subfigure}%
    \begin{subfigure}{0.5\linewidth}
        \centering
        \begin{tikzpicture}
            \tikzstyle{ann} = [font=\small,inner sep=1pt]
            \node[inner sep=0pt] (mesh) at (0.48,0)
                {\includegraphics[width=0.65\textwidth]{mesh_example.pdf}};
            \draw[arrows=->] (0,0) to (2,0);
            \draw[arrows=->] (0,0) to (0,2.75);
            \node[ann] at (-0.2, 2.55) {$z$};
            \node[ann] at (1.8, -0.2) {$\rho$};
            \node[ann] at (-0.5, 2.85) {(b)};
        \end{tikzpicture}
    \end{subfigure}

    \caption{(a) Simulation domain $\Omega$ is a truncation of the full $\rho\geq 0$ half-plane.
    We can impose both Dirichlet and Neumann boundary conditions at will on different
    sections ($\Gamma_D,\Gamma_N$) of the domain boundary $\partial\Omega$.
    (b) Triangular mesh of the domain $\Omega$
    supports arbitrary spatial refinement.}
\label{fig:simulation_domain}
\end{figure}

The above interaction operators conserve the magnetic quantum number $m$, and by default,
we simulate only the $m=0$ subspace. Correspondingly, the coordinate space domain is a two-dimensional slice of
the cylindrical coordinates, $\Omega_{\infty} = \left\{ \rho \geq 0,z\in\mathbb{R}\right\}$, which
we truncate for numerical simulations
to a \emph{finite} domain $\Omega$ [see Fig.~\ref{fig:simulation_domain}(a)].

\subsection{Time independent Schrödinger equation}

\noindent On the domain $\Omega_\text{TI}$ the time-independent Schrödinger equation (TISE)
can be written as
\begin{equation}
\label{eqn:tise}
        -\frac{1}{2\rho}\frac{\partial}{\partial\rho} \left(\rho \frac{\partial\psi_k}{\partial \rho} \right)
                                   -\frac{1}{2}\frac{\partial^2 \psi_k}{\partial z^2} +V(\rho, z) \psi_k= E_k \psi_k,
\end{equation}
where $E_k, \psi_k$ is the $k$th eigenpair. This equation is accompanied by continuity boundary condition at $\rho=0$,
\begin{equation}
    \lim\limits_{\rho\to0+}\left(\rho\frac{\partial \psi}{\partial \rho} \right) = 0.
\end{equation}
Elsewhere on the boundary $\partial \Omega_\text{TI}$ we can choose freely between
zero Neumann boundary conditions (ZNBCs) and zero Dirichlet boundary conditions (ZDBCs)
-- depending on the requirements of our model.
For example, in Fig.~\ref{fig:simulation_domain}(a) where we have imposed ZNBC on
the upper boundary of the simulation domain, $\Gamma_N$, and ZDBC on the arc $\Gamma_D$.

To discretize TISE, we begin by looking for eigenpairs $E_k\in\mathbb{R}, \psi_k\in F$
of the weak form corresponding to Eq.~\eqref{eqn:tise} and the boundary conditions described above.
The weak form is given by
\begin{equation}
\label{eqn:tise_weak}
\begin{split}
    &-\frac{1}{2}\int\limits_{\Omega_\text{TI}} \sum\limits_{\alpha=\rho,z} \frac{\partial \chi}{\partial\alpha} \frac{\partial\psi_k}{\partial\alpha}\rho\,\mathrm{d}\rho \,\mathrm{d}z    \\
           & + \int\limits_{\Omega_\text{TI}} \chi V(\rho, z) \psi_k \rho\,\mathrm{d}\rho\,\mathrm{d}z \\
           & = E_k \int\limits_{\Omega_\text{TI}} \chi \psi_k
           \rho\,\mathrm{d}\rho\,\mathrm{d}z\; \forall \chi \in  F,
\end{split}
\end{equation}
where $F=\left\{ \psi\in H^1(\Omega_\text{TI})\, |\, \psi(\Gamma_D) = 0 \right\}$ is
the standard Sobolev space on $\Omega_\text{TI}$ of
real-valued  $L^2$-integrable functions with $L^2$ integrable first derivatives. Notice also that $F$ includes ZDBCs, and
the natural inner product is
\begin{equation}
\label{eqn:inner_product}
\braket{\chi}{\psi} = \int\limits_{\Omega_\text{TI}} \chi(\rho, z)^* \psi(\rho,z)\,\rho\,\mathrm{d}\rho\,\mathrm{d}z.
\end{equation}

Furthermore, we must restrict ourselves to a finite dimensional
approximation of $F$. First, the simulation domain $\Omega_\text{TI}$ is described using
an unstructured triangular mesh which supports arbitrary refinement
[see, e.g., Fig.~\ref{fig:simulation_domain}(b)].
Next, we construct a basis of continuous low-order Lagrange polynomials $\phi_i$,
each with compact support on the mesh elements. This basis spans a finite dimensional
function space
\begin{equation}
    F_h = \mathrm{span}\left\{ \phi_i \right\}_{i=0}^{N-1} \subset F
\end{equation}
where the TISE weak form [Eq.~\eqref{eqn:tise_weak}] can be written
as a finite dimensional generalized Hermitian eigenvalue equation
\begin{equation}
\label{eqn:tise_discrete}
    (\mathbf{T} + \mathbf{V})\bm \psi_k = \mathbf{S} E_k \bm \psi_k.
\end{equation}
Here $\bm \psi_k$ is a vector of the real-valued expansion coefficients of the
$k$th eigenstate, $E_k$ the corresponding eigenvalue, and
\begin{align}
S_{ij} &=  \bra{\phi_i}\ket{\phi_j},\label{eqn:overlap_matrix}\\
T_{ij} &= -\frac{1}{2}\sum\limits_{\alpha=\rho,z} \bra{\frac{\partial\phi_i}{\partial\alpha}}\ket{\frac{\partial\phi_j}{\partial\alpha}},\text{ and}\label{eqn:kinetic_matrix}\\
V_{ij} &= \braket{\phi}{V(\rho,z)\phi_j}\label{eqn:potential_matrix}
\end{align}
are the overlap, kinetic energy, and potential energy matrices, respectively.
Note that since the static potential of the system is implemented only via Eq.~\eqref{eqn:potential_matrix}
which involves integration, we can use any static potential for which Eq.~\eqref{eqn:potential_matrix} is finite, e.g., the Coulomb potential.

\subsection{Time dependent Schrödinger equation}
\noindent In a similar manner as with TISE, TDSE
can be discretized as
\begin{equation}
        \iu \mathbf S \dot{\bm \psi}(t) = \left( \mathbf T + \mathbf V + \mathbf W
    \right) \bm\psi(t) \label{eqn:discrete_tdse},
\end{equation}
where $\bm\psi(t)$ is a vector of the \emph{complex-valued} expansion coefficients $c_i(t)$ of the wave function $\psi(\rho,z,t) = \sum c_i(t) \phi_i(\rho,z)$.
$\bm S$, $\bm T$, and $\bm V$ are constructed the same way as for TISE (except we add an imaginary absorbing potential to $\bm V$),
and the light-matter interaction matrix is given by
\begin{equation}
\label{eqn:interaction_matrix}
W_{ij} = \mel**{\phi_i}{W_s\left(\rho, z,\frac{\partial}{\partial \rho}, \frac{\partial}{\partial z}, t\right)}{\phi_j}
\end{equation}

The basis of the discrete function space for TDSE can, in general, be different from the
one used for TISE. This is useful in practice since often
the stationary states can be computed in a much smaller simulation domain
than needed for an accurate description of the TDSE. We can change the function space
by three methods: (1) using a larger simulation domain for the TDSE,
(2) refining the mesh according to the requirements of the TDSE simulation,
and (3) change the degree of our basis functions according to the required accuracy.

For evolving the discretized state, i.e., the expansion coefficients $\bm \psi(t)$ according to Eq.~\eqref{eqn:discrete_tdse},
\fiend\ implements two approximations of the time evolution operator:
the Crank-Nicolson (CN) method~\cite{cn_operator} and
the generalized-$\alpha$ method~\cite{alpha_algorithm}.

The Crank-Nicolson method can be written as
\begin{equation}
\begin{split}
\psi(t+\Delta t) = &\left[\mathbf{S} + \frac{ \iu \Delta t }{2}\mathbf{H}\left(t+\frac{\Delta t}{2}\right) \right]^{-1} \\
                           &\left[\mathbf{S} - \frac{ \iu \Delta t
                           }{2}\mathbf{H}\left(t+\frac{\Delta t}{2}\right)
                           \right]  \psi(t) + \mathcal{O}(\Delta t^2),
\end{split}
\end{equation}
and it is a unitary time-reversible transformation~\cite{cn_operator} as long
as the matrix inversion is performed to a sufficient accuracy.

The generalized-$\alpha$ method was originally developed for integrating the equations of motion arising from
fluid dynamics~\cite{alpha_algorithm}, but we have successfully applied it to the TDSE integration.
To the best of our knowledge,
the $\alpha$-method has not been proven to be unitary nor is it exactly time-reversible.
Nevertheless, we have achieved accuracies comparable to the CN method -- and in the case of extremely inhomogeneous
vector potentials, the generalized-$\alpha$ method seems to be significantly more stable than CN.

\subsection{Incorporation of boundary conditions}
\label{sec:boundary_conditions}

\noindent The continuity boundary condition at $\rho=0$ and ZNBCs are automatically included by
dropping the boundary integrals when deriving Eqs.~\eqref{eqn:tise_weak} and~\eqref{eqn:kinetic_matrix}.
However, the ZDBCs must be included in the system matrices via the following modifications:
the rows and columns of $\bm S$, $\bm T$, $\bm V$, and $\bm W$ corresponding
to the degrees of freedom (DOFs) on the boundary $\Gamma_D$ with ZDBCs set to zero.
Only for $\bm S$ we must set the corresponding diagonal elements to one to ensure invertibility.
Note that it's crucial to zero out not only the rows but also the corresponding columns of the matrices
to retain hermiticity.

Embedding ZNBCs via the vanishing boundary terms when deriving Eqs.~\eqref{eqn:tise_weak} and~\eqref{eqn:kinetic_matrix}
causes a practical issue: Some operators such as $\hat{p}_z=-\iu\partial_z$, $\hat{p}_\rho = -\iu \left(\partial_\rho+\frac{1}{2\rho}\right)$,
and $\bm A \cdot \bm {\hat{p}} + \bm {\hat{p}} \cdot \bm A$
become \emph{non-Hermitian}. Minor adjustments at the boundaries can remedy this, namely,
\begin{align}
    \hat{p}_\rho  \to -\iu\left[\partial_\rho+\frac{1}{2\rho}-\frac{1}{2}\delta(\bm r - \Gamma_N )n_\rho(\bm r) \right],\\
    \hat{p}_z \to -\iu\left[\partial_z-\frac{1}{2}\delta(\bm r - \Gamma_N )n_z(\bm r) \right],\text{ and}\\
    \frac{1}{2}(\bm A \cdot \bm {\hat{p}}  + \bm {\hat{p}} \cdot \bm A ) \to \frac{1}{2}\left(\bm A \cdot \bm {\hat{p}} + \bm {\hat{p}} \cdot \bm A \right)\nonumber \\
    \phantom{\to\,}-\frac{\iu}{2} \bm{A}\cdot\bm n(\bm r) \delta(\bm r - \Gamma_N ),
\end{align}
where $\bm n(\bm r) = [n_\rho(\bm r), n_z(\bm r)]^T$ is the outwards facing unit normal on the simulation domain boundary at point $\bm r$.
This trick has been previously used to obtain the correct Hermitian operators in hyperspherical coordinates~\cite{radial_momentum_operator}.

Another aspect to consider is the weak enforcement of ZNBCs compared to ZDBCs. If there was
significant electron density on the Neumann boundary $\Gamma_N$, it will start to oscillate and eventually violate the ZNBC.
This can be remedied when operating with matrix inverses as in the second step of the CN-propagator. We can add an extra error term, e.g., of the form
\begin{equation}
\gamma\sum\limits_{i} \left\vert\, \int\limits_{\Gamma_N} \chi^*_i(\rho, z) \nabla\psi(\rho,z,t) \cdot \D\bm S \right\vert^2\!\!, \gamma\in\mathbb R
\end{equation}
to the error estimator in our numerical implementation. This will remove -- or at least lessen -- the issue arising from the weak enforcement of ZNBCs.

\section{Implementation}
\label{sec:implementation}
\subsection{Overview}

\noindent We have implemented finite element discretization of TISE and TDSE in the Python software package
\fiend\ following the recipes of Sec.~\ref{sec:theory}. \fiend\ is written in Python 3.6 and it utilizes
reliable libraries commonly pre-installed in high-performance clusters and supercomputers. For
sparse linear algebra we use the Portable, Extensible Toolkit for Scientific Computation
(\pkg{PETSc})~\cite{petsc1,petsc2,petsc4py}
and the Scalable Library for Eigenvalue Problem Computations (\pkg{SLEPc})~\cite{slepc1,slepc2}.
Meshing and other standard FEM-related parts are built on top of the components of
the FEniCS project~\cite{fenics1,fenics2,dolfin1,ffc1,ffc2,ufl1,fiat1}, and filesystem IO
is largely based on \pkg{HDF5}~\cite{HDF5} via \pkg{h5py}~\cite{h5py1,h5py2}. In postprocessing
and visualization we use also \pkg{numpy}~\cite{numpy},
\pkg{scipy}~\cite{scipy}, and \pkg{matplotlib}~\cite{matplotlib}. \fiend\ is parallelized using \pkg{MPI} via
\pkg{mpi4py}~\cite{mpi4py1,mpi4py2,mpi4py3}.

The \pkg{PETSc}-dependency is slightly complicated as \pkg{PETSc} needs to be compiled
either with real number support (x)or with complex number support. \fiend, on the other hand,
needs real numbers for TISE and post-processing but complex numbers for propagation.
Consequently, the user must install \emph{both} the real and complex versions of \pkg{PETSc}/\pkg{SLEPc} stacks
and switch between these for different simulation steps.
This factitious requirement for two different installations of \texttt{PETSc}
can be revisited in the future upon the completion of \texttt{DOLFIN-X}~\cite{dolfinx} and
\texttt{FFC-X}~\cite{ffcx} including the merge of complex number support in \fenics~\cite{baratta_gsoc},

To ease installation, we provide a Docker~\cite{docker} container with \fiend\ and all its dependencies
preinstalled, see \url{https://hub.docker.com/r/solanpaa/fiend/} for more details.
In the same spirit, \fiend\
is also available in the Python Package Index~\cite{fiend_pypi} and can be installed with \pkg{pip}~\cite{pip}.

\tikzset{
    state/.style={
           rectangle,
           rounded corners,
           draw=black, very thick,
           minimum height=2em,
           inner sep=2pt,
           text centered,
           },
}
\begin{figure}
\centering
\begin{tikzpicture}[->,>=stealth']
\node (BEGIN) {};

\node[state,
anchor=north,
below of=BEGIN,
node distance = 3.0cm] (TISE)
 {\begin{tabular}{l}
     \textbf{1. Solve TISE}\\\\
  \parbox{0.9\linewidth}{\begin{itemize}
   \item Mesh $\Omega_{TI}$
   \item Assemble matrices
   \item Solve eigenproblem
   \item Save results
  \end{itemize}
  }
 \end{tabular}};

 \node[state,
  below of=TISE,
  node distance=3.0cm,
  anchor=north] (TDSE)
 {\begin{tabular}{l}
     \textbf{2. Preparation step}\\
  \parbox{0.9\linewidth}{\begin{itemize}
   \item Mesh $\Omega_{TD}$
   \item Remesh stationary states if necessary
   \item Assemble matrices
   \item Save matrices and states and convert them to complex numbers
  \end{itemize}
  }
 \end{tabular}};

  \node[state,
  below of=TDSE,
  node distance=3.5cm,
  anchor=north] (PROPAGATION)
 {\begin{tabular}{l}
  \textbf{3. Propagation}\\
  \parbox{0.9\linewidth}{\begin{itemize}
   \item Load matrices and states from step 2
   \item Construct time-dependent Hamiltonian
   \item Compute $\bm \psi(t)$ and save data
  \end{itemize}}
 \end{tabular}};
    \node[state,
  below of=PROPAGATION,
  node distance=3.0cm,
  anchor=north] (ANALYSIS)
 {\begin{tabular}{l}
  \textbf{4. Analysis}\\
  \parbox{0.9\linewidth}{\begin{itemize}
   \item Post-processing
   \item Multiple example scripts available
  \end{itemize}
  }
 \end{tabular}};

 \path (BEGIN)         edge[line width = 1pt]  node[anchor=south,left]{Load \fenics/\pkg{PETSc}}
                                    node[anchor=north,right]{with real numbers} (TISE);
 \path (TISE)   edge[line width = 1pt]  node[anchor=south,right]{}
                                    node[anchor=north,below]{} (TDSE);
 \path (TDSE)   edge[line width = 1pt]  node[anchor=south,left]{Switch to \pkg{PETSc}}
                                    node[anchor=north,right]{with complex numbers} (PROPAGATION);
 \path (PROPAGATION)    edge[line width = 1pt]  node[anchor=south,left]{Switch to \pkg{PETSc}}
                                    node[anchor=north,right]{with real numbers} (ANALYSIS);\end{tikzpicture}
\caption{Typical simulation structure with \fiend\ together with the required changes to the runtime environment between the steps.}
\label{fig:workflow}
\end{figure}
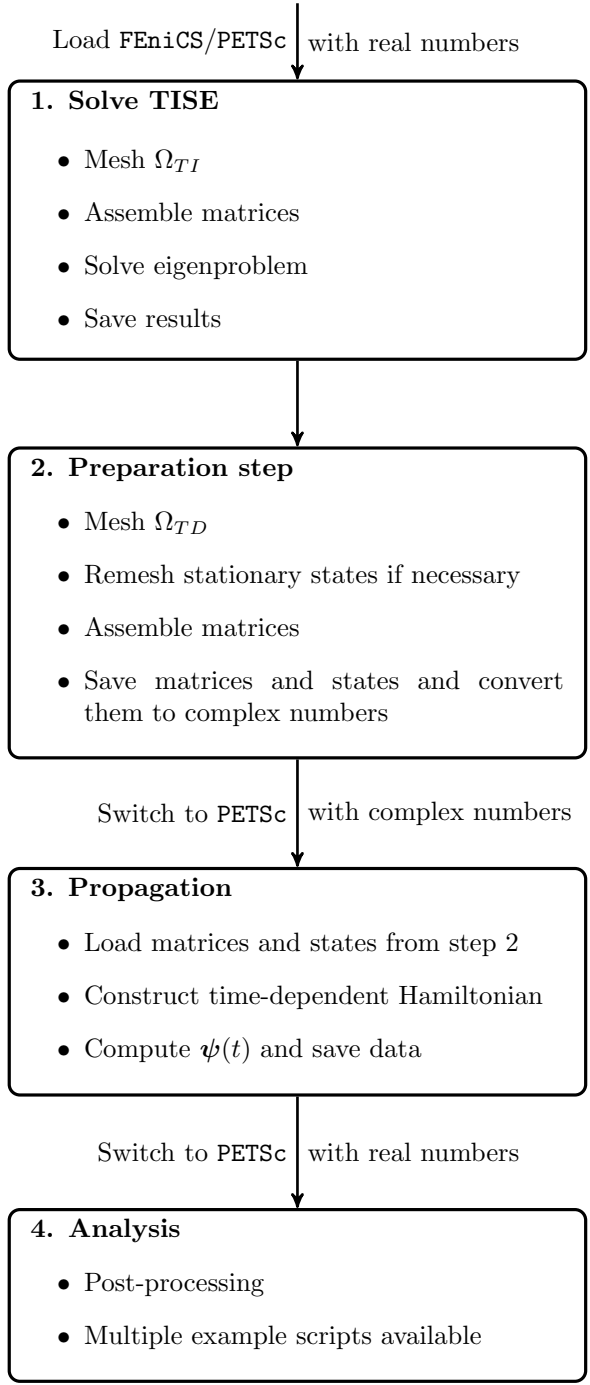


\subsection{Library usage and numerical methods}
\noindent A typical usage of the \fiend\ suite is demonstrated in Fig.~\ref{fig:workflow}.
First we solve the TISE to obtain a set of stationary states.
This includes meshing the domain for the time independent simulation and assembling the system
matrices~\eqref{eqn:overlap_matrix}--\eqref{eqn:potential_matrix} with appropriate boundary conditions
(see Sec.~\ref{sec:boundary_conditions}).

The TISE eigenproblem, Eq.~\eqref{eqn:tise_discrete},
is solved (by default) with Rayleigh Quotient Conjugate Gradient (RQCG) method
combined with classical Gram-Schmidt orthogonalization of the Krylov subspace basis
with adaptive iterative refinement for increased numerical stability. RQCG is a variational method
which essentially minimizes the Rayleigh quotient
\begin{equation}
\frac{\bm \psi_k^\dagger \bm H_0 \bm \psi_k}{\bm \psi_k^\dagger \bm S \bm \psi_k}
\end{equation}
of a desired number of orthonormal vectors with respect to the bilinear product induced by $\bm S$~\cite{slepc2}.
Upon convergence, this corresponds to $k$ smallest real eigenpairs of TISE.

Step 2 (see Fig.~\ref{fig:workflow}) is to prepare the discrete
description of the TDSE. This includes meshing the domain for the time dependent problem,
interpolating stationary states to the new mesh, and also assembly of the system matrices
~\eqref{eqn:overlap_matrix}--\eqref{eqn:potential_matrix} and the time-independent part(s) of Eq.~\eqref{eqn:interaction_matrix}.

The propagation in Step. 3 (Fig.~\ref{fig:workflow}) requires the user to load
the environment with complex number \texttt{PETSc}. The time-dependent Hamiltonian is
constructed from the matrices computed in the previous step, and the discrete TDSE~\eqref{eqn:discrete_tdse}
is solved with \texttt{PETSc}'s propagators. We also provide code templates for easy implementation
of new propagators.

Finally, we provide a set of example scripts for postprocessing. These
scripts include, e.g., visualization of the integrated electron density in time and
temporal shape of the laser field, but there are also scripts to compute more complex observables such
as the high-harmonic and photoelectron spectra.

All the numerical methods in \fiend\ depend on efficient sparse linear algebra operations implemented
in \texttt{PETSc}. By default, matrix inversions and the solution of linear equations are carried out with
Generalized Minimal RESidual method (\mbox{GMRES})~\cite{gmres}, where the convergence criterion is computed
with respect to the norm induced by the inner product in Eq.~\eqref{eqn:inner_product}.
Unfortunately, the \pkg{PETSc} linear algebra backend does not allow us
to build the basis of the Krylov subspace with respect to a custom inner product.
However, we have found the basis built with the standard Hermitian inner product to be
sufficient performance-wise. Direct solvers such as \texttt{SuperLU\_DIST}~\cite{superlu1,superlu2,superlu3}
and \texttt{MUMPS}~\cite{mumps1,mumps2} are supported only with matrices that can be explicitly
constructed without too high cost.

\subsection{Note on meshing}
\label{sec:mesh_refinement}
\noindent We provide a way to generate meshes with arbitrary refinement in the coordinate space.
The user should supply a function returning the maximum allowed cell circumradius\footnote{Radius of the smallest circle enclosing the given triangular cell.}
at coordinate $\bm r$, and the mesh can be refined until all cells of the mesh have circumradius below the desired one.

By default, the maximum cell circumradius is given by~\cite{PhysRevB.52.R2229}
\begin{equation}
\begin{split}
\max\text{CR}(\bm r) = \text{CR}_\text{asymp}\left[1-\left(1-\frac{\text{CR}_\text{ref}}{\text{CR}_\text{asymp}}\right)\right.\\
\times\left.\frac{R_\text{ref}}{r}\tanh\left(\frac{r}{R_\text{ref}}\right)\exp\left(-\frac{r^2}{R_\text{trans}^2}\right)\right].
\end{split}
\end{equation}
For $r<R_\text{ref}$ the cell circumradii are below  $\text{CR}_\text{ref}$, and as $r\to R_\text{trans}$, the maximum cell circumradius increases
monotonically to $\text{CR}_\text{asymp}$.

\section{Examples and test cases}

\label{sec:demos}

\subsection{Field-free propagation}

\noindent To assess the stability and accuracy of the numerical methods implemented in \fiend,
we compute a field-free propagation of an electron prepared in a superposition of Hydrogen 1s and 2s states,
\begin{equation}
        \psi(t = 0) = \frac{1}{\sqrt{2}} \left( \psi_{1s} + \psi_{2s} \right).
\end{equation}
This state is propagated up to \SI{100}{\au} of time. The simulation domain up to $R=$~\SI{30}{\au} is meshed using the default refinement introduced in Sec.~\ref{sec:mesh_refinement}
with $R_\text{ref}=$~\SI{4}{\au}, $R_\text{ref}=$~\SI{10}{\au}, $\text{CR}_\text{ref}=$~\SI{0.01}{\au} and $\text{CR}_\text{asymp}=$~\SI{0.5}{\au}
We assess the accuracy of the numerical solution by comparing the simulated state to the exact result:
\begin{equation}
\left\vert 1 - \vert \bra{\psi_{\text{exact}}}\ket{\psi_{\text{\fiend}}} \vert^2 \right\vert.
\end{equation}
The stability is investigated by evaluating how much the norm of the state differs from unity, i.e.,
\begin{equation}
\left\vert 1 - \Vert \psi_{\text{\fiend}} \Vert^2\right\vert.
\end{equation}

Both the $\alpha$-propagator and the CN propagator reach an accuracy of 0.00524 \% for the overlap with the time-step \SI{0.05}{\au},
and the norm of the simulated state deviates from unity only by $3.22\cdot10^{-5}\,\%$.
This demonstrates the applicability of the propagators for the simulation of FE-discretized
TDSE in cylindrical coordinates.

\subsection{High-harmonic generation}

\noindent Next we demonstrate the applicability of \fiend\ to simulate high-order
harmonic generation (HHG). According to the three-step model~\cite{threestepmodel1,threestepmodel2}, an atomic electron is excited
to continuum and driven to oscillate around the atom~\cite{strong_field_laser_physics_principles}. Upon recombination with
the parent ion, the excess energy is released as high-energy photons.

Full 3D simulation of HHG via TDSE
is a difficult task due to the long extent of the electron wave function when driven
by a short laser pulse. In order to obtain a decent description of the single-atom
response for HHG, a large simulation domain is needed.
We setup a simulation domain of radius \SI{800}{\au} with a more refined mesh
at the origin and gradually sparser mesh for larger radii.

\begin{figure}[t]
\includegraphics[width=\linewidth]{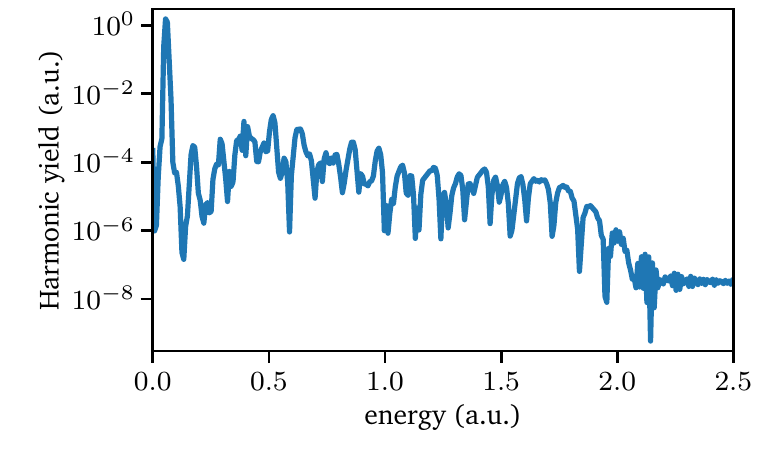}
\caption{High-order harmonic spectrum of a hydrogen atom under a few-cycle femtosecond laser pulse.}
\label{fig:hhg}
\end{figure}

The intensity spectrum of the single-atom response is proportional to the square of the Fourier transform
of the dipole acceleration~\cite{harmonic_spectrum},
\begin{equation}
S(\omega) \propto \left\vert \expval{\ddot{D}_z}(\omega)\right\vert^2,
\end{equation}
where we compute the dipole acceleration via Ehrenfest's theorem~\cite{dipole_acceleration_ehrenfest}, i.e.,
\begin{equation}
\expval{\ddot{D}_z}(t) = \mel**{\psi(t)}{-\frac{\partial V}{\partial z}}{\psi(t)}.
\end{equation}
In Fig.~\ref{fig:hhg} we show the computed high-order harmonic spectrum
computed for a hydrogen atom under a laser pulse with carrier frequency corresponding to \SI{800}{\nano\meter},
intensity full width at half maximum \SI{4.8}{\femto\second}, and electric field peak intensity \SI[per-mode=repeated-symbol]{36}{\giga\volt\per\meter}.
The spectrum is a typical HHG spectrum from a femtosecond laser pulse, and it
ends at the cut-off energy $\sim$ \SI{2}{\au} in a decent agreement with
the three-step model's cutoff $\sim$ \SI{1.7}{\au}~\cite{threestepmodel1,threestepmodel2}.

\subsection{Metal nanotips and inhomogeneous fields}

\begin{figure*}[t]
\includegraphics[width=\textwidth]{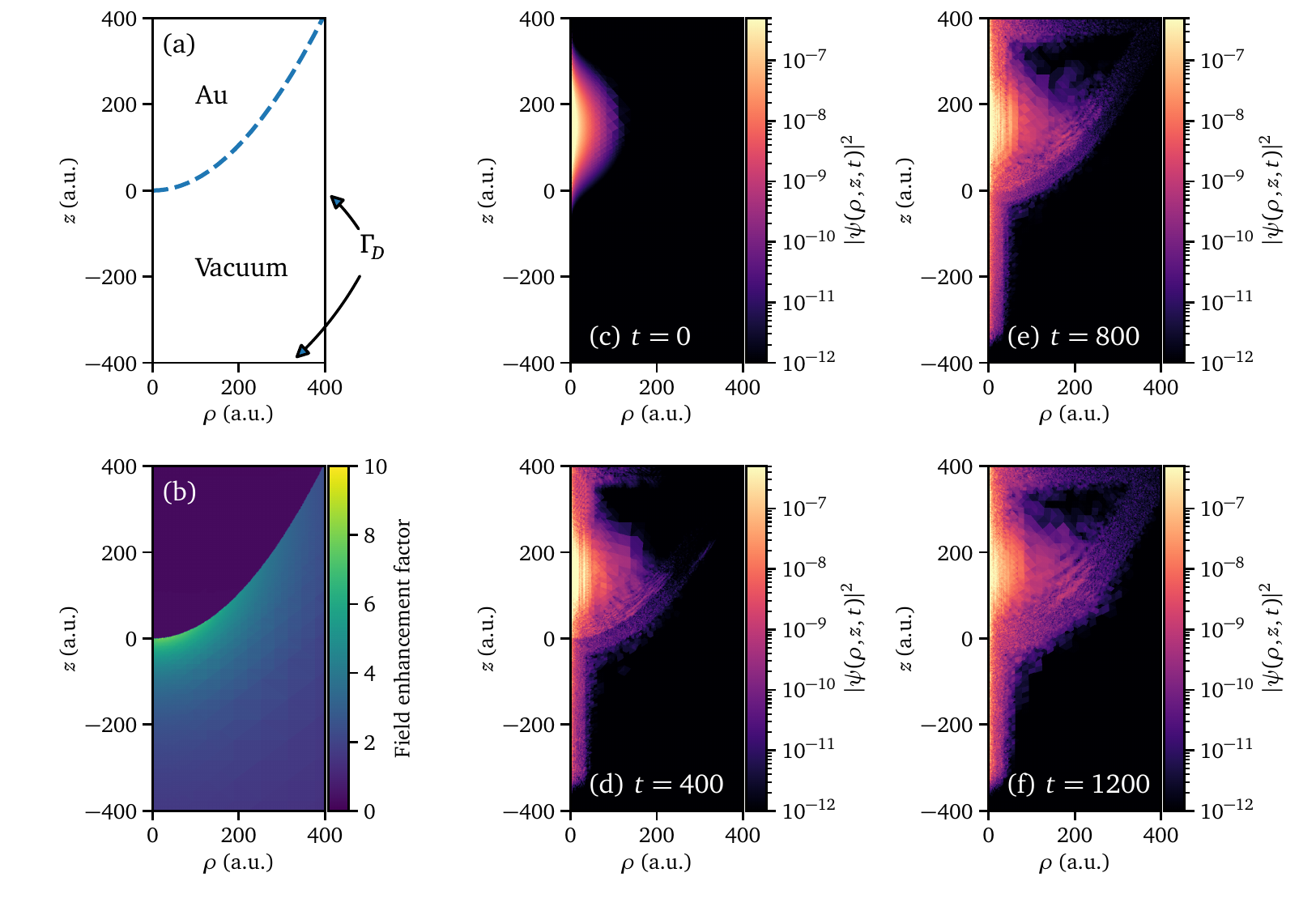}
\caption{(a) Setup of the nanotip geometry where we have a sharp boundary between the gold nanotip and vacuum.
(b) Computed spatial enhancement profile of the electric near field in a quasistatic approximation. (c)-(f) Snapshots of the electron
density during time evolution.}
\label{fig:nanotip}
\end{figure*}

\noindent Metal nanotips have recently attracted attention due to their ability to
enhance the laser electric field at the tip apex via plasmonic
effects (see, e.g.,~\cite{Piglosiewicz2013,kruger2012,Wimmer2014,doi:10.1063/1.4934681,doi:10.1063/1.4991681,1367-2630-9-5-142,herink2012,PhysRevLett.105.147601,Forg2016,field_local_resc}).
\fiend\ is capable of computing single-electron
dynamics of these metal nanotapers including correctly the interaction with
the \emph{inhomogeneous} plasmon enhanced laser field.

The first step is to compute the plasmon enhanced near field. Here we consider
a gold nanotip with apex radius \SI{7}{\nano\meter} and full opening angle of $20\,{}^\circ$ as
demonstrated in Fig.~\ref{fig:nanotip}(a). For typical laser wavelengths, such as
\SI{800}{\nano\meter} used here, a quasistatic description of the laser vector potential
is applicable, i.e.,
\begin{equation}
    \mathbf A(\mathbf r, t) = \mathbf A_s(\mathbf r) f(t),
\end{equation}
where the spatial form $\mathbf A_s$ is the same as for the electric field
\begin{equation}
    \mathbf A_s(\mathbf r) = \mathbf E(\mathbf r) = -\nabla U(\mathbf r).
\end{equation}
The electrostatic potential $U(\mathbf r)$ can be computed from the
Poisson equation
\begin{equation}
    -\nabla \cdot \left[ \epsilon(\mathbf r) \nabla U(\mathbf r) \right] = 0,
    \label{eqn:poisson}
\end{equation}
where $\epsilon(\mathbf r)$ is the dielectric function of the material at
position $\mathbf r$. Inside the gold nanotip the complex dielectric
function (at \SI{800}{nm}) is
$\epsilon_{\mathrm{Au}}=-24.061 + 1.5068\, \iu$~\cite{gold_dielectric_function,rii} and at vacuum
$\epsilon_{\mathrm{vac}}=1$.

We can describe Eq.~\eqref{eqn:poisson}
using cylindrical coordinates. The interface condition
\begin{equation}
\epsilon_{\mathrm{Au}}\nabla u\vert_{\bm r \to \Gamma_\text{tip-vac}}-\epsilon_{\mathrm{vac}}\nabla u\vert_{\bm r\to\Gamma_\text{vac-tip}}=0
\end{equation}
gives rise to the weak form
\begin{equation}
\begin{split}\label{eqn:poisson}
    &\epsilon_{\mathrm{vac}} \!\!\!\!\!\int\limits_{\mathrm{vacuum}} \!\!\!\!\!\nabla v^*(\mathbf r)
    \cdot \nabla u(\mathbf r)\, \rho\,\mathrm{d}\rho\,\mathrm{d}z \\
     +\, &\epsilon_{\mathrm{Au}} \int\limits_{\mathrm{tip}} \nabla v^*(\mathbf r)
    \cdot \nabla u(\mathbf r)\,\rho \,\mathrm{d}\rho\,\mathrm{d}z = 0~\forall
    v\in \hat{U}
\end{split}
\end{equation}
with the test function space
\begin{equation}
\begin{split}
\hat{U} = \big\{v: \Omega\to\mathbb{C} \mid
\Re(v),\,\Im(v)\in H^1(\Omega) : \\v(\Gamma_D) = 0 \big\}
\end{split}
\end{equation}
and the trial function space
\begin{equation}
\begin{split}
U = \big\{ u: \Omega\to\mathbb{C} \mid \Re(u),\,\Im(u)\in H^1(\Omega):\\
 u(\mathbf r\in\Gamma_D) = z \big\}.
\end{split}
\end{equation}
The simulation domain is a rectangular area, and the ZDBCs described above
are imposed on the external boundaries denoted by $\Gamma_D$. Note that
Fig.~\ref{fig:nanotip}(a) demonstrates the domain and boundaries for the quantum
simulation -- for Poisson equation we employ a much larger simulation domain. The
equation~\eqref{eqn:poisson} can be solved easily with the \texttt{FEniCS}
FEM-suite and we provide an example script at \texttt{demos/nanotip/1\_near\_field.py}

The computed spatial distribution of the field, $\Vert A_s(x, y, z) \Vert$, is demonstrated in
Fig.~\ref{fig:nanotip}(b). The field is at its maximum at the nanotip apex
and decays rapidly with the distance to the tip. We note that this quasistatic solution
provides field-enhancement factor of $f\approx10$ which is comparable with experimental
results~\cite{nanotip_enhancement}.

In TDSE, we use a potential well for the static potential,
\begin{align}
    U(\mathbf r) = \left\{\begin{array}{rl}
        0,  & \text{vacuum}\\
        -\Phi, & \text{inside the tip,}
    \end{array}\right.
\end{align}
where $\Phi=\SI{5.31}{eV}$ is the work function of Au (111)-surface~\cite{au111_work_function}.

We prepare the electron as a Gaussian wave packet [Fig.~\ref{fig:nanotip}(c)] and assemble the system matrices
-- \emph{including the inhomogeneous field} -- with
\texttt{demos/nanotip/3\_prepare\_tdse.py}. The preparation step is slightly more
complex than for linearly polarized vector potentials, but the provided example script
should work as a template for further expansions.

We use a \SI{800}{\nano\meter} pulse with \SI{15}{\femto\second} full width at
half maximum for the Gaussian envelope. The maximum electric field amplitude without
nanostructure enhancement is set to \SI{30}{\giga\volt/\nano\meter}.
Snapshots of the electron density during propagation are shown in
Figs.~\ref{fig:nanotip}(c)-(f). As expected, the electron emission is concentrated at the
tip apex where the field enhancement is at its highest. Only a small
percentage ($\sim$\num{2}\,\%) of the electron density is absorbed at the
simulation box boundary.

\section{Summary}
\label{sec:summary}
\noindent We have presented \fiend\ --
a versatile solver for single-particle quantum dynamics in cylindrically symmetric systems.
This Python package provides an easy path for the study of nonlinear strong-field phenomena
in atoms and nanostructures under both homogeneous and \emph{inhomogeneous} vector potentials.
Moreover, the finite element discretization used by \fiend\ can adapt to complicated system geometries.
The package is parallelized using \pkg{MPI}, and much of the high-performance computing
is delegated to state-of-the-art numerical libraries. \fiend\ is modular and easy to extend,
and we provide comprehensive documentation for the program code.

We have demonstrated the capabilities of \fiend\ by simulating two different scenarios.
The simulation of high-order harmonic generation from a laser-driven hydrogen atom
shows that \fiend\ can be applied to traditional strong-field phenomena in atoms.
Furthermore, a simulation of the photoionization of a gold nanotip demonstrates \fiend's suitability
for studying plasmon-enriched strong-field phenomena in nanostructures.

\section*{Acknowledgements}
\noindent We are grateful to Matti Molkkari and Joonas Keski-Rahkonen for insightful discussions.
 This work was supported by the Alfred Kordelin foundation and
 the Academy of Finland (grant no. 304458).
 We also acknowledge CSC – the Finnish IT Center for Science – for computational resources.

\bibliographystyle{elsarticle-num}
\bibliography{refs}

\end{document}